\documentstyle[12pt,graphicx]{article}
\title{RELIC GRAVITATIONAL  WAVES AND CMB
POLARIZATION IN THE ACCELERATING  UNIVERSE}

\author{Y. ZHANG*,
        W. ZHAO, T. Y. XIA, X. Z. ER, and
        H. X. MIAO \\
        Center for Astrophysics,\\
University of Science and Technology of China \\
         Hefei, Anhui, 230026  China  \\
           *yzh@ustc.edu.cn}

\begin{document}
\maketitle
\baselineskip=21truept


\newcommand{\be}{\begin{equation}}
\newcommand{\ee}{\end{equation}}
\newcommand{\bee}{\begin{eqnarray}}
\newcommand{\een}{\end{eqnarray}}
\newcommand{\ba}{\begin{eqnarray}}
\newcommand{\ea}{\end{eqnarray}}

\sf
\begin{abstract}
In this paper we briefly present our works on the relic
gravitational waves (RGW) and the CMB polarization in the
accelerating universe.
The spectrum of RGW has been obtained,
showing the influence of the dark energy.
Compared with those  from non-accelerating  models,
the shape of the spectrum is approximately similar,
nevertheless, the amplitude of RGW now
acquires a suppressing factor of the ratio of matter over dark
energy $\propto \Omega_m/\Omega_{\Lambda}\sim 0.4$ over almost the
whole range of frequencies.
The RGW spectrum is then used as the source
to calculate the spectra of CMB polarization.
By a two half Gaussian function as an approximation to
the visibility function during the photon decoupling,
both the ``electric" and ``magnetic" spectra
have been analytically derived, which are quite
close to the numerical ones.
Several physical elements that
affect the spectra have been examined, such as the decoupling
process, the inflation, the dark energy, the baryons, etc.
\end{abstract}



\section{Introduction}

The existence of gravitational waves is
a major prediction of General Relativity that has not yet been
directly detected.
On other hand,
inflationary models  predict,
among other things,
 a stochastic  background of relic gravitational waves
(RGW) generated during the very early stage of expanding universe.
\cite{starobinsky}$^-$\cite{Grishchuk77}
Therefore, the detection of RGW
plays a double role in relativity and cosmology.
For a number of gravitational  detections,
ongoing or under development,
the spectrum of RGW represents one of their major scientific goals.
However, the current expansion of the universe has been found to be
an accelerating one, probably driven by dark energy.
This will have important implications on RGW and its detections.
As is known, the cosmic background radiation
has certain degree  of polarization generated via Thompson scattering
during the decoupling in the early universe.
\cite{Basko}$^-$\cite{BondEfstathiou87}
In particular, if the tensorial perturbations (RGW)
are present at the photon decoupling in the universe,
then magnetic type of polarization will
be produced.\cite{Polnarev80}$^-$\cite{HuWhite}
This would be a characteristic feature of RGW on very large scales,
since the density perturbations will not
generate this magnetic type of polarization.
Besides the generation of linear polarization,
the rotation of linearly polarized EM propagation by RGW
has also been first studied in Refs.\cite{Ni,Ni2},
and WMAP polarization data has already been used to constrain
the effect in cosmological distance to 0.1 rad,
which is important for fundamental physics.
For both theoretical and observational studies,
it is necessary to examine
the effects of the dark energy on RGW and on CMB
anisotropies and polarization.
In this talk I shall present our calculational results
on these issues.

First I will present briefly our result
of the spectrum of RGW,
both analytical and numerical,
in the
accelerating Universe $\Omega_{\Lambda}+\Omega_m = 1$.
As a double check,
we have also derived an approximation of the spectrum
analytically.
The results
from both calculations are consistent with each other.
Discussions are given on the possible detections.

Then I will mention sketchily our analytic
calculation  of the CMB polarization
produced by Thompson scattering in the presence of the RGW.
The resulting spectra  are quite
close to the numerical one computed from the CMBFAST code,
and have several improvements over the previous analytic results.
Moreover, the formulae bear the explicit dependence
on such important processes,
as the decoupling, the inflation, the dark energy, the baryons.

\section{ RGW in the Accelerating Universe}

Consider a spatially flat Universe
with the  Robertson-Walker metric
\be
ds^2=a^2(\tau) [  d\tau^2-(\delta_{ij}+h_{ij})dx^idx^j ],
\ee
where $h_{ij}$ is $3\times 3$ symmetric,
representing the perturbations,
$\tau$ is the conformal time.
The scalar factor $a(\tau)$ is given for the following
various stages.
The initial stage (inflationary)
\be \label{i} a(\tau ) = l_0 \mid
{\tau} \mid ^{1+\beta},  \,\,\,\,\, -\infty < \tau \leq  \tau_1,
\ee
where $1+\beta<0$, and $\tau_1<0$. The special case of
$\beta=-2$ is the de Sitter expansion of inflation.
The  reheating stage
\be
a(\tau) = a_z(\tau-  \tau_p)^{1+\beta_s},
\,\,\,\,\,  \tau_1 \leq  \tau \leq  \tau_s,
\ee
allowing
a general reheating epoch.\cite{grish1,grish01}
The radiation-dominated stage
\be a(\tau) = a_e(\tau  -\tau_e),
\,\,\,\,\, \tau_s \leq  \tau \leq  \tau_2.
\ee
The matter-dominated stage
\be a(\tau) =  a_m(\tau  -\tau_m)^2 ,
\,\,\,\,\,  \tau_2  \leq  \tau \leq  \tau_E,
\ee
where $\tau_E$ is
the time when the dark energy density $\rho_{\Lambda}$ is equal to
the matter energy density $\rho_m$.
The  redshift $z_E$ at the time $\tau_E$ is given by  $1+z_E =
(\frac{\Omega_{\Lambda}}{\Omega_m})^{1/3}$.
If the current values
$\Omega_{\Lambda} \sim  0.7$ and $\Omega_m \sim  0.3$ are taken,
then  $1+z_E \sim 1.33$.
The accelerating stage
(up to the present time $\tau_H$)\cite{zh}$^-$\cite{zh06}
\be \label{acc}
a(\tau) =  l_H |\tau-  \tau_a| ^{-\gamma}, \,\,\,\,\,
 \tau_E \leq  \tau  \leq \tau_H ,
\ee
where  $\gamma$ is a parameter.
For  the de Sitter acceleration
with $\Omega_{\Lambda}=  1$ and $\Omega_m=0$, one has  $\gamma=1.0$.
We have numerically solved the Friedman equation
\be
\left(\frac{a'}{a^2}\right)^2= H^2 (\Omega_{\Lambda}+\Omega_m a^{-3})
\ee
with $a'\equiv da(\tau)/d\tau$,
and have found that
the expression of (\ref{acc}) gives a good fitting
with
$\gamma = 1.05$  for  $\Omega_{\Lambda}=0.7$,
$\gamma = 1.06$  for $\Omega_{\Lambda} = 0.65$,
$\gamma = 1.048$ for $\Omega_{\Lambda} = 0.75$,
and $\gamma = 1.042$ for $\Omega_{\Lambda} = 0.80$.
\cite{zh}$^-$\cite{zh06}

There are ten constants in the above expressions of $a(\tau)$,
except  $\beta$ and $\beta_s$,
that are imposed upon as the model parameters.
By the continuity conditions  of $a(\tau)$ and  $a(\tau)'$ at the  four
given joining points $\tau_1$, $\tau_s$, $\tau_2$, and $\tau_E$,
one can fix only eight constants.
The other two  constants can
be fixed by the overall normalization of $a$ and by the observed
Hubble constant as the expansion rate. Specifically,
we put $a(\tau_H)=l_H$ as the normalization, i.e.
\be \label{c}
|\tau_H  -   \tau_a| = 1,
\ee
and the constant $l_H  $ is fixed by the following calculation
\be
\frac{1}{H} \equiv  \left(\frac{a^2}{a'}   \right)_{\tau_H}  =
\frac{l_H}{\gamma}  .
\ee
To completely fix the joining conditions
we  need to specify the time instants
 $\tau_{1}$, $\tau_{2}$,  $\tau_s$, and $ \tau_{E}$.
From the consideration of physics of the Universe,
we take the following specifications\cite{zh}$^-$\cite{zh06}:
$a(\tau_H)/a(\tau_E)=1.33$,
$a(\tau_E)/a(\tau_2)=3454$,
$a(\tau_2)/a(\tau_s)=10^{24}$,
and $a(\tau_s)/a(\tau_1)=300$.
The physical wavelength $ \lambda $ is related to
the comoving wave number $k$ by
\be
 \lambda \equiv \frac{2\pi a(\tau)}{k}.
 \ee
The wave number corresponding to
the present Hubble radius is
$  k_H = 2\pi a(\tau_H )/l_H  =2\pi $.

The gravitational wave field is
the tensorial portion of $h_{ij}$, which is transverse-traceless
$\partial_i h^{ij}=0$,  $\delta^{ij}h_{ij}=0$,
and the wave equation is
\be
\partial_{\mu}(\sqrt{-g}\partial^{\mu}h_{ij}({\bf{x}} ,\tau))=0 .
\ee
For  a fixed wave vector $\bf k$ and a fixed polarization state
$\sigma=+$ or $\times$, the wave equation reduces to
\be \label{h}
 h_k^{(\sigma)''}+2\frac{a'}{a}h_k^{(\sigma)'} +k^2h^{(\sigma)} _k
  =0.
\ee
Since the equation of
 $h_{\bf k}^{(\sigma)} (\tau) $
 for each polarization $\sigma$ is the same,
 we denote $h_{\bf k}^{(\sigma)}
(\tau) $ by $h_{\bf k}(\tau) $ in the following.
Once the mode function  $h_k(\tau)$ is known,
the spectrum $h(k,\tau)$ of RGW is given by
\be   \label{spectrum}
h(k,\tau) = \frac{4l_{\rm Pl}}{\sqrt{\pi}}k|h_k(\tau)|,
\ee
and the spectral  energy density  $\Omega_g(k) $ of the
GW is defined
\be \label{omega}
 \Omega_g(k) = \frac{\pi^2}{3}h^2(k, \tau_H)
 \left(\frac{k}{k_H} \right)^2   ,
 \ee
which is dimensionless.

The initial conditions of RGW
are taken to be during the inflationary stage.
For a given wave number $k$, the corresponding wave  crossed over the
horizon at a time  $\tau_i$,
i.e. when the wave length was equal to the Hubble radius:
$\lambda_i = 2\pi a(\tau_i)/k$  to $1/H(\tau_i)$.
Now
the initial condition is taken to be
\be \label{inv}
h(k, \tau_i) = A \left(\frac{k}{k_H}\right)^{2+\beta} ,
\ee
where the constant $A$ is to be fixed by the CMB anisotropies.
The power spectrum for
the primordial scalar perturbations
is $P_s(k)\propto |h( k, \tau_H)|^2 $,
and its spectral index $n_s$ is defined as
$P(k) \propto  k^{n_s-1}$.
Thus one reads off the relation $n_s = 2\beta +5$.
The exact de Sitter  expansion of $\beta = -2$ leads to $n_s=1$,
yielding the so-called the scale-invariant primordial spectrum.

Any calculation of the spectrum of  RGW  has to
fixed the normalization of the amplitude.
One can  use the CMB anisotropies to constrain the amplitude,
receiving the contributions from
both the  scalar perturbations  and the RGW.
The ratio is defined as
\be\label{ratio}
r= P_h/P_s,
\ee
the value of which has not been observationally fixed up yet.
Here the ratio $r$ is taken as a parameter.
This will  determine the overall factor $A$  in (\ref{inv}).
Using the observed  CMB anisotropies\cite{spergel,spergel2}
 $\Delta T/T \simeq 0.37 \times 10^{-5}$ at $l \sim 2$,
one has
\be \label{initialcd}
h(k_H,
\tau_H) =  0.37  \times 10^{-5}r.
\ee
Then the spectrum $h(k, \tau_H)$ at the present time $\tau_H$ is fixed.

\begin{figure}[pt]
\centerline{\includegraphics[width=12cm]{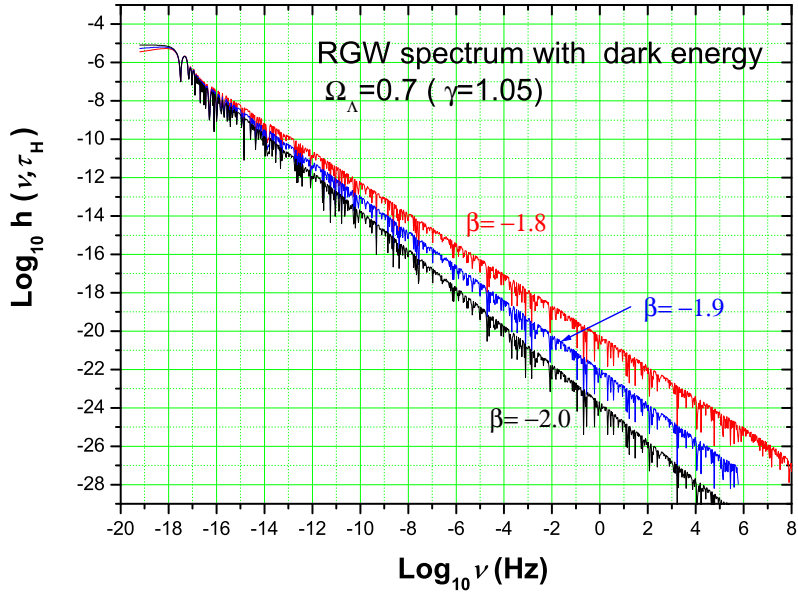}}
\caption{\label{amplitude-105}
For a fixed $\gamma=1.05$,
the exact spectrum $h(\nu,\tau_H)$ is plotted for
three inflationary models of $\beta=-1.8,-1.9,-2.0$, respectively. }
\end{figure}

\begin{figure}
\centerline{\includegraphics[width=12cm]{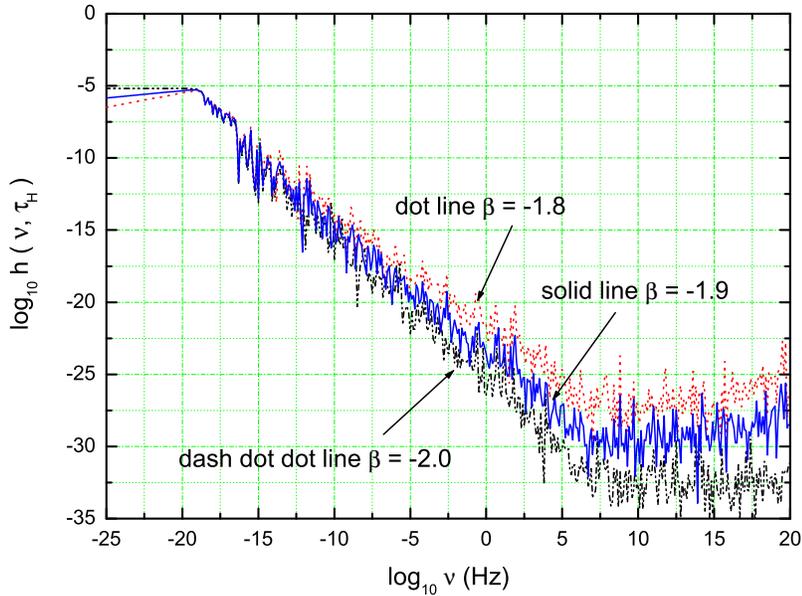}}
\caption{ \label{numerical-spectrum}
The numerical spectra $h(k,\tau_H)$ in the accelerating universe
for $\beta=-1.8, -1.9$ and $-2.0$, respectively. }
\end{figure}

\begin{figure}
\centerline{\includegraphics[width=12cm]{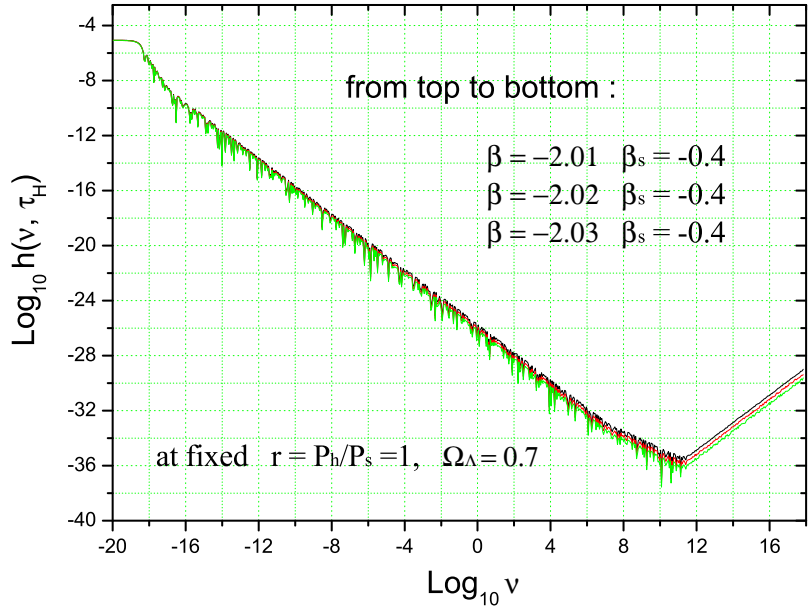}}
\caption{ \label{-2spectrum}
The exact spectra $h(k,\tau_H)$ in the accelerating universe
for $\beta=-2.01, -2.02$ and $-2.03$, respectively. }
\end{figure}

Writing the mode function $h_k(\tau)=\mu_k(\tau)/a(\tau)$
in Eq. (\ref{h}),  the equation for $\mu_k(\tau)$  becomes
\be \label{mu}
\mu''_k+  \left(k^2-\frac{a''}{a}\right)\mu_k=0.
\ee
 For a scale factor of  power-law form
 $  a(\tau )   \propto \tau^{\alpha}$,
 the general exact solution is
 \[
\mu_k(\tau) = c_1(k\tau)^{\frac{1}{2}}J_{\alpha-\frac{1}{2}}(k\tau)
+ c_2(k\tau) ^{\frac{1}{2}}J_{\frac{1}{2}-\alpha }(k\tau),
 \]
 where the constant $c_1$ and $c_2$ are to determined by
continuity of the function $\mu_k(\tau)$ and the  time
derivative $(\mu_k(\tau)/a(\tau))'$ at the  instances
joining two consecutive  stages.
We have analytically solved  the equation for the various stages,
from the inflationary  through the accelerating stage.
The final expressions are lengthy and we do not write
 down them here. \cite{zh}$^-$\cite{zh06}
However, the resulting spectrum will plotted for illustration.
Taking the  ratio  $r=0.37$ and $\gamma=1.05$,
we have plotted the exact spectrum $h(k,\tau_H)$
in Fig. \ref{amplitude-105}
for three inflationary models with
$\beta=-1.8, -1.9$, and $ -2.0$,
and $\beta_s =  0.598$,    $ -0.552$, and $-0.689$,
respectively.
We also plot the spectrum from the numerical calculation in
Fig. \ref{numerical-spectrum}.
And Fig. \ref{-2spectrum} shows the spectra for $\beta<-2.0$.
\begin{figure}
\centerline{\includegraphics[width=12cm]{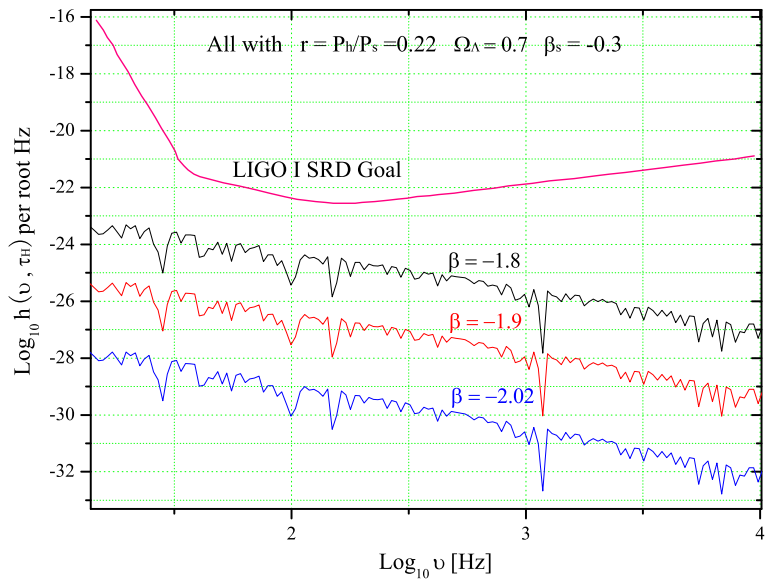}}
\caption{ \label{ligo-105}
For $\gamma=1.05$
the root mean square spectrum $h(\nu,\tau_H)/\sqrt{\nu}$ is plotted
for the models of $\beta=-1.8,-1.9,-2.02$ to compare  with the
sensitivity curve from  S5 of LIGO.\cite{ligo} }
\end{figure}
\begin{figure}
\centerline{\includegraphics[width=12cm]{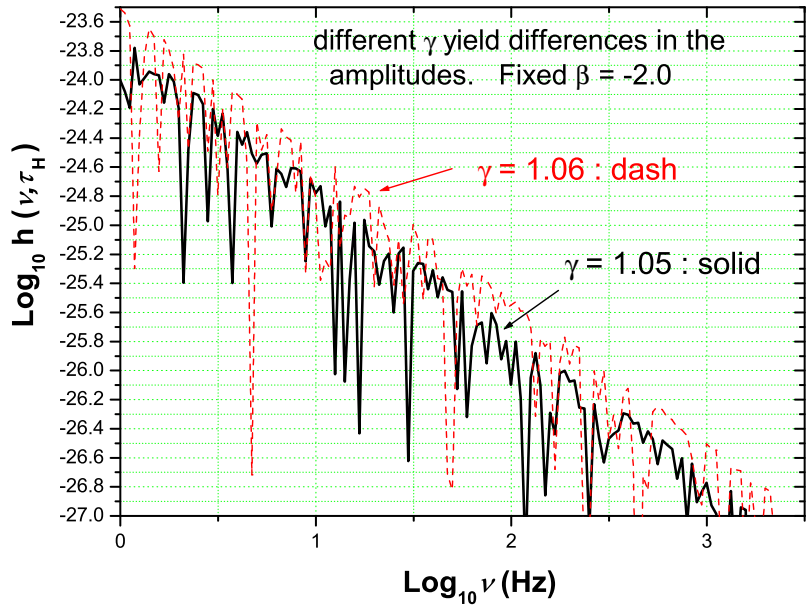}}
\caption{\label{fine20}
The amplitude of $h(\nu,\tau_H)$ for the model $\gamma=1.06$ is $\sim 50 \%$
higher than that of model $\gamma=1.05$. }
\end{figure}
\begin{figure}
\centerline{\includegraphics[width=12cm]{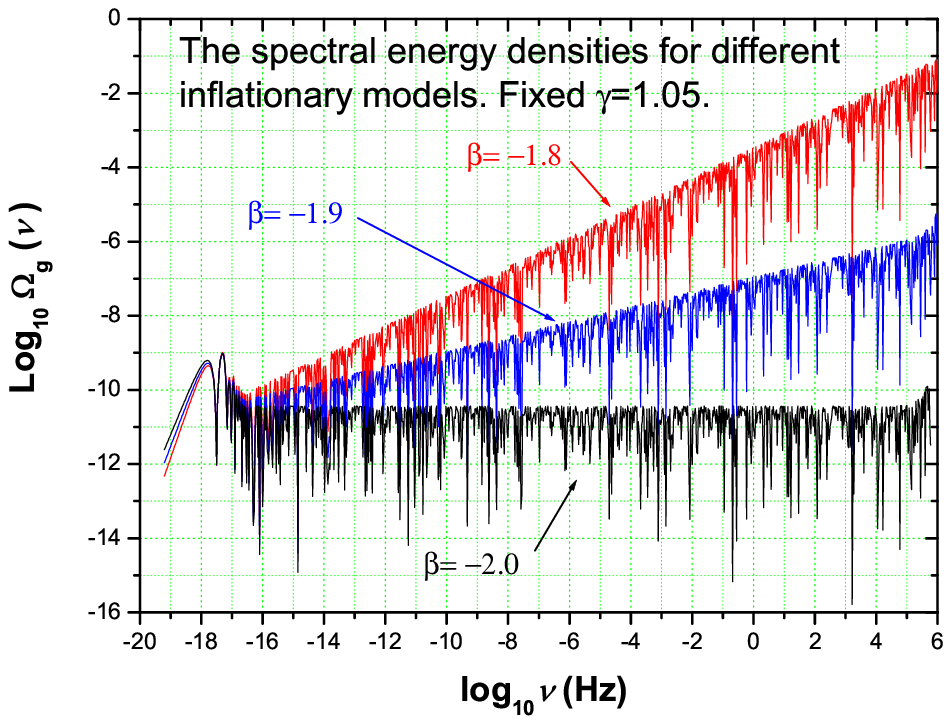}}
\caption{\label{energy105}
The spectral energy density $\Omega_g(\nu)$ is plotted for the models of
$\beta=-1.8$, $\beta=-1.9$, and $\beta=-2.0$.
The model $\beta= -1.8$ is ruled out by BBN constraint.}
\end{figure}
Fig. \ref{ligo-105} compares the root mean square spectrum
  $h(\nu, \tau_H)/\sqrt{\nu}$
of the model  $\gamma=1.05$ with the sensitivity of LIGO I SRD
\cite{ligo}$^-$\cite{BAbbbott05}
in the  frequency range  $\nu = 10 \sim10^4$Hz.

The spectrum $h(\nu, \tau_H)$ depends
on the dark energy $\Omega_{\Lambda}$ through
the parameter $\gamma$.
In Fig. \ref{fine20}
we have  plotted the spectra in a narrow range of frequencies.
It is seen that the amplitude in the model $\gamma=1.06$ is about
$\sim 50 \%$ greater than that in the model $\gamma=1.05$.
That is, in the accelerating Universe with $\Omega_{\Lambda} =0.65$
the amplitude of relic GW is $\sim 50 \%$ higher than
the one with $\Omega_{\Lambda} =0.7$.
This  difference is probably difficult
to detect at present.
However, in principle, it does provide
a new way to tell the dark energy fraction
$\Omega_{\Lambda}$  in the Universe.

Let us examine the spectral energy density
$\Omega_g(\nu)$ and its constraints.
Fig. \ref{energy105} is the the plots of
$\Omega_g(\nu)$ defined in Eq. (\ref{omega}) for $\gamma=1.05$.
If we use the result LIGO third science run\cite{BAbbbottprl,BAbbbott05}
of the energy density bound for the flat spectrum with
$\Omega_0 <8.4\times 10^{-4}$ in the $69-156$ Hz band,
then the model $\beta=-1.8$ is ruled out,
but the models $\beta \leq-1.9$ survive.
However, this LIGO constraint is not as stringent as
the constraint by
the so-called nucleosynthesis bound.
\cite{maggiore}$^-$\cite{maia2}
\be \label{energylimit}
\int \Omega_g(\nu) \, d(\log \nu) \leq  0.56\times 10^{-5}.
\ee
Note that this is bound on the total GW energy density integrated over
all frequencies.
The integrand function should also have a bound
$ \Omega_g(\nu) < 0.56\times 10^{-5}$
in the interval of frequencies $\delta (\log \nu) \simeq 1$.
By this constraint it is also seen from Fig. \ref{energy105}
that only the model $\beta = -2.0$ are still robust.

We have also obtained  the following
expressions for the analytic approximate spectrum
\be \label{ini}
h(k, \tau_H) = A \left(\frac{k}{k_H}\right)^{2+\beta}, \,\,\,\,\, k\leq k_E;
\ee
\be \label{eh}
h(k,\tau_H)\approx
   A \left(\frac{k}{k_H}\right)^{\beta-1}\frac{1}{(1+z_E)^{3+\epsilon}},\,\,\,\,\,
   k_E \leq k \leq k_H;
\ee
\be \label{h2}
h(k,\tau_H) \approx
   A    \left(\frac{k}{k_H}\right)^\beta   \frac{1}{(1+z_E)^{3+\epsilon}},\,\,\,\,\,
   k_H \leq k \leq k_2;
\ee
\be \label{2s}
h(k,\tau_H)
  \approx  A  \left(\frac{k}{k_H} \right)^{\beta+1}  \frac{k_H}{k_2}
  \frac{1}{(1+z_E)^{3+\epsilon}},
  \,\,\,\,\,\,\, k_2 \leq k \leq k_s;
\ee
\be \label{fin}
  h(k,\tau_H) \approx
 A   \left(\frac{k_s}{k_H}\right)^{\beta_s}\frac{k_H}{k_2}
    \left(\frac{k}{k_H}\right)^{\beta-\beta_s+1} \frac{1}{(1+z_E)^{3+\epsilon}},
     \,\,\,\,\,\,\, k_s \leq k \leq k_1,
\ee
where the small parameter $\epsilon \equiv
(1+\beta)(1-\gamma)/\gamma$.
Approximately
$\frac{1}{(1+z_E)^{3+\epsilon}}
\sim \frac{1}{(1+z_E)^3}
=\Omega_m/\Omega_{\Lambda}$.
This extra factor reflects the effects of acceleration
caused by the dark energy.
Some of other works on RGW can be found in
Refs.\cite{Tashiro}$-$\cite{WangZhang08},
and the effects of neutrino free-streaming have been
recently computed in Ref.\cite{weinberg}$-$\cite{MiaoZhang}.

\section{CMB Polarization}

At the beginning,  I mentioned that
the magnetic polarization of CMB gives another way to
detect RGW.
During the era prior to the decoupling in the early Universe,
the Thompson scattering of anisotropic radiation by
 free electrons can
give rise to the linear polarization only,
so we only consider
the polarized distribution function of photons $f=(I_l, I_r, U)$
whose components are associated with the Stokes parameters:
 $I= I_l+I_r$ and $Q=I_l-I_r$.
The evolution of the photon distribution function is given by
the Boltzmann equation\cite{chan}
\be \label{boltzeq}
 \frac{\partial f}{\partial\tau}+\hat{n}^i
\frac{\partial f}{\partial x^i}= -\frac{d
\nu}{d\tau}\frac{\partial f}{\partial \nu}
-q(f-J),
 \ee
where $\hat{n}^i$ is the unit vector in the direction
$(\theta,\phi)$ of photon propagation, $q$ is the
differential optical depth and has the meaning of scattering rate.
The scattering term $q(f-J)$
describes the effect of the
Thompson scattering by free electrons, and the term
$-\frac{d\nu}{d\tau}\frac{\partial f}{\partial \nu}$
reflects the effect of
variation of frequency due to the metric perturbations
through  the Sachs-Wolfe formula
\be \label{sachs}
\frac{1}{\nu}\frac{d\nu}{d\tau}
= \frac{1}{2} \frac{\partial
  h_{ij} }{\partial \tau}\hat{n}^i \hat{n}^j.
\ee
In the presence of perturbations $h_{ij}$,
either scalar  or  tensorial,
the distribution function  will be perturbed and can be
written as
 \be
 f(\theta,\phi)= f_0\left[
 \left(
 \begin{array}{c}
 1\\
 1\\
 0\\
 \end{array}
 \right) +f_1\right] ,
 \ee
where  $f_1$  represents the perturbed portion,
$  f_0(\nu)$
is the usual blackbody distribution.
The tensorial type perturbations $h_{ij}$,
representing the RGW,
has two independent, $+$ and $\times$,  polarization.
\[
h_{ij} = h^{+}_{ij}+ h^{\times}_{ij} = h^{+} \epsilon^+_{ij}
           + h^{\times}\epsilon^{\times}_{ij}.
 \]
To simplify  the Boltzmann equation (\ref{boltzeq}),
for the $h_{ij} = h^{+} \epsilon^{+}_{ij}$ polarization,
one writes $f_1$ in the form\cite{Basko,Polnarev85}
 \be \label{f1+}
 f_1= \frac{\zeta }{2} \left(1-\mu^2\right)\cos2\phi
 \left( \begin{array}{c}
1\\
1\\
0
\end{array}
\right)
 + \frac{\beta }{2} \left(
\begin{array}{c}
(1+\mu^2)\cos2\phi\\
-(1+\mu^2)\cos2\phi\\
4\mu\sin2\phi
\end{array}
\right),
 \ee
where $\zeta\propto I_l +I_r = I$
 represents the anisotropies of photon  distribution,
and $\beta \propto  I_l -I_r =Q$ represents the
polarization of photons.
From the Boltzmann equation,
 upon  taking Fourier transformation,
retaining only the terms linear in  $h_{ij}$,
and performing the integration over $d\mu$,
one arrives at a set of two
equations for
the $+$ polarization, \cite{Polnarev85,Ng,zhz}
\begin{eqnarray} \label{eqn:rewritten1}
   \dot{\xi}_k+\left[ik\mu+q\right]\xi_k  =
      \frac{d \ln f_0 }{d \ln \nu_0}    \dot{h}^+_k ,
\end{eqnarray}
\begin{eqnarray}
     \dot{\beta}_k+\left[ik\mu+q\right]\beta_k  =  \frac{3q}{16}
     \int^{1}_{-1}
     d\mu'\left[\left(1+\mu'^2\right)^2\beta_k
     -\frac{1}{2}\left(1-\mu'^2\right)^2\xi_k\right].
     \label{eqn:rewritten}
\end{eqnarray}
where $\xi_k  \equiv \zeta_k +\beta_k $,
the over dot $``\cdot"$ denotes $d/d\tau$.
For the blackbody spectrum $f(\nu_0)$ in the
Rayleigh-Jeans zone  one has $\frac{d\ln f_0(\nu_0)}{d\ln\nu_0}\approx 1$.
The equations are the same for  the $\times$ polarization.
In the following
we simply  omit the sub-index $k$ of wavenumber,
the GW polarization notation, $+$ or $\times$,
since both $h^+$ and $h^\times$ are similar in computations.
In general, it is difficult to give the
exact solution of
$\beta$ and $\xi$, but once derived,
they can be expanded in terms of
the Legendre functions
\[
\xi(\mu) =\sum_l (2l+1) \xi_l P_l(\mu),  ~ ~~\,\,
\beta(\mu) = \sum_l (2l+1) \beta_l P_l(\mu),
\]
with the Legendre components
\be\label{xil}
\xi_l(\tau) = \frac{1}{2}\int_{-1}^1 \,d\mu\, \xi(\tau,\mu)P_l(\mu),
~ ~~\, \,
\beta_l(\tau) =\frac{1}{2}\int_{-1}^1\, d\mu\, \beta(\tau,\mu)P_l(\mu).
\ee
It can be shown that
the spectrum for electric type polarization is given by
\begin{equation}
     C_l^{\rm GG}=\frac{1}{16\pi}\int \,
 \left|\frac{(l+2)(l+1)\beta_{l-2}}{(2l-1)(2l+1)}+\frac{6
     (l-1)(l+2)\beta_{l}} {(2l+3)(2l-1)} +
     \frac{l(l-1)\beta_{l+2}}{(2l+3)(2l+1)}\right|^2\,k^2dk,\label{gg}
\end{equation}
the spectrum for magnetic  type polarization is given by
 \be
 C_l^{\rm CC}=\frac{1}{4\pi}\int\,
   \left|\frac{(l+2)\beta_{l-1}}{2l+1}+\frac{(l-1)\beta_{l+1}}
  {2l+1}\right|^2\,k^2dk . \label{cc}
\ee

\begin{figure}[pt]
\centerline{\includegraphics[width=12cm]{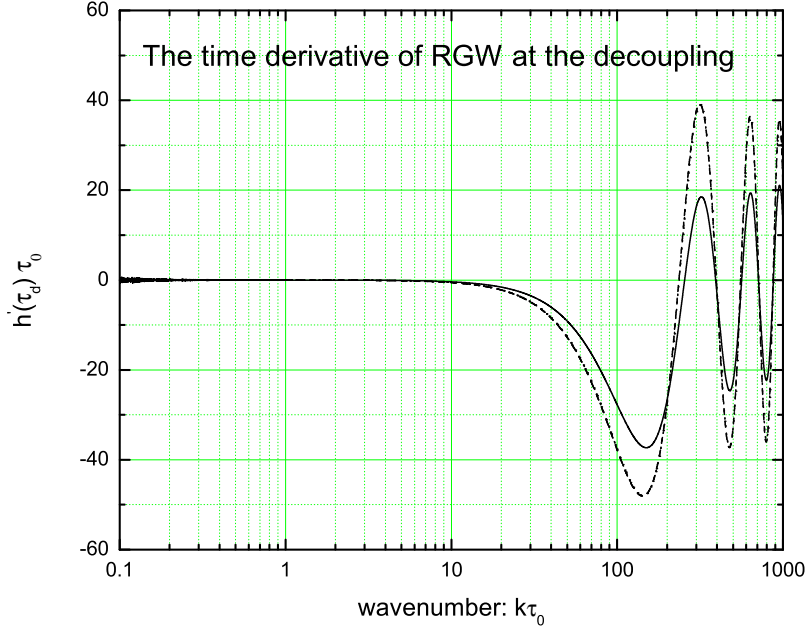}}
\caption{\label{hdot}
The derivative of RGWs $\dot h(\eta_d)$ as a function of $k$.
The solid line is the
sudden transition approximation, the dash line is
that of the WKB approximation,
which is nearly overlapped with the dot line of the numerical result. }
\end{figure}

As Eq. (\ref{eqn:rewritten1})  shows,
one needs the time derivative of  $\dot h(\tau)$
to solve for $\xi$ and $\beta$.
For both polarization,  $+, \times$,
the wave equation of the relic GW  has been given
in Eq. (\ref{h}).
The initial condition is taken to be
 \be \label{hi}
 h(\tau=0)=h(k),~~\dot{h}(\tau=0)=0,
 \ee
with the  primordial power spectrum
 \be \label{powerh}
 \frac{k^3}{2\pi^2}  |h(k)|^2  =P_h(k)
 =A_T \left(\frac{k}{k_0}\right)^{n_T},
 \ee
where  $A_T$ is the amplitude,
$k_0= 0.05$ Mpc$^{-1}$ is the pivot wavenumber,
and $n_T$ is the the tensor spectrum index. Inflationary models
generically predicts $n_T\approx0$, a nearly scale-invariant
spectrum.
The resulting  $\dot{h}(\tau_d)$
is plotted  in   Fig. \ref{hdot}.

One solves the ionization equations during the
recombination to give the
differential optical depth $q$.
Then one obtains the visibility function $V(\tau)$,
\begin{equation}
\label{V}
V(\tau)=q(\tau)e^{-\kappa(\tau_0,\tau)},
\end{equation}
which satisfies $ \int_{0}^{\tau_0}V(\tau)d\tau=1$
and describes the probability that a given
photon last scattered at  time $\tau$,
where the optical depth function $\kappa(\tau_0,\tau)$ is related to
$q(\tau)$ by $q(\tau) =-d\kappa(\tau_0,\tau)/d \tau $.
\cite{peebles}$^-$\cite{Hu}
Fig.
\ref{V} shows the profile of $V(\tau)$
by the the numerical result from the CMBFAST,
which  is sharply peaked around the last scattering.
In calculation  it is usually fitted by
a Gaussian form\cite{zalda}\cite{prit}
\begin{equation} \label{v}
V(\tau)=V(\tau_d) \exp\left(-\frac{(\tau-\tau_d)^2}{2
\Delta\tau_d^2}\right),
\end{equation}
where $\tau_d$ is the the decoupling time, and $\Delta\tau_d$
is the thickness of decoupling.
The  WMAP data\cite{spergel} gives $\Delta\tau_d/\tau_0=0.00143$.
Then,  taking $V(\tau_d)\tau_0=279$ in (\ref{v})
yields a fitting shown in Fig. \ref{V},
which has large errors, compared with the numerical one.
To improve the fitting  of $V(\tau)$,
we take the following analytic expressions,
consisting of two half-Gaussian functions,\cite{zhaozhang083006}
\begin{equation}\label{halfgaussian1}
V(\tau)=V(\tau_d) \exp\left(-\frac{(\tau-\tau_d)^2}{2
\Delta\tau_{d1}^2}\right),~~~(\tau<\tau_d);
\end{equation}
\begin{equation}\label{halfgaussian2}
V(\tau)=V(\tau_d) \exp\left(-\frac{(\tau-\tau_d)^2}{2
\Delta\tau_{d2}^2}\right),~~~(\tau>\tau_d);
\end{equation}
with $\Delta\tau_{d1}/\tau_0=0.00110$,
$\Delta\tau_{d2}/\tau_0=0.00176$, and
$(\Delta\tau_{d1}+\Delta\tau_{d2})/2=\Delta\tau_{d}$.
Fig. \ref{V}  shows that the half-Gaussian model fits
the numerical one much better than the Gaussian fitting.
This difference  will subsequently cause
a variation in the polarization spectra.
\begin{figure}[pt]
\centerline{\includegraphics[width=12cm]{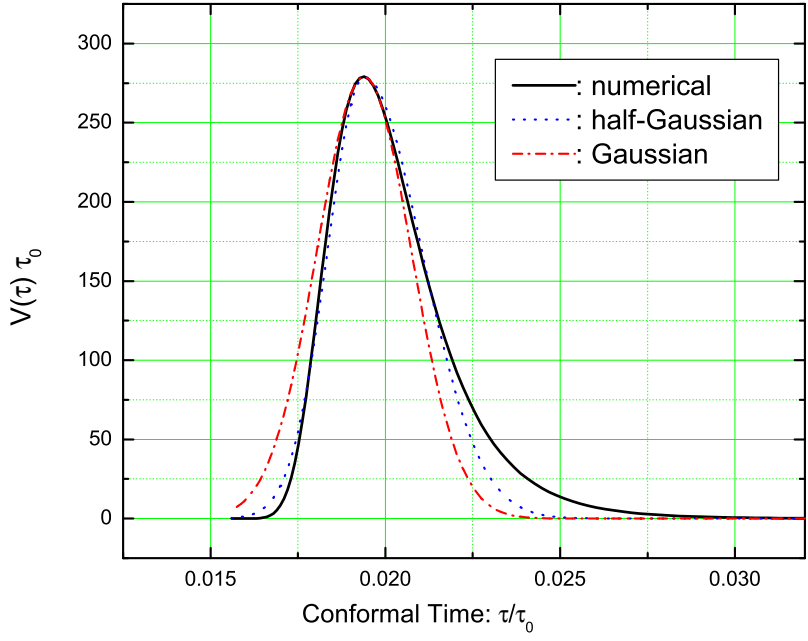}}
 \caption{\label{V}
The visibility function $V(\tau)$ around the decoupling.
The half-Gaussian model improves
the Gaussian model by  $\sim 11.5 \%$.}
 \end{figure}

We now look for an approximate and analytic solution of
Eqs. (\ref{eqn:rewritten1}) and (\ref{eqn:rewritten}).
On smaller scales the photon diffusion
will cause some damping in the anisotropy and polarization.
Taking care of this effect  to the second order
of  the small parameter $1/q \ll 1$,
by some analysis
one arrives at the expression for the component of the polarization
\be \label{bl}
 \beta_l(\tau_0)=\frac{1}{10}  \Delta\tau_d\, i^l
 \int_0^{\tau_0}d\tau V(\tau) \dot{h}(\tau)   j_l(k(\tau-\tau_0))
 \int_1^{\infty}\frac{dx}{x}e^{-\frac{3}{10}\kappa(\tau)
x}e^{-\frac{7}{10}\kappa(\tau)},
\ee
where $\kappa(\tau) \equiv \kappa(\tau_0,\tau)$,
$x \equiv \kappa(\tau')/\kappa(\tau)$,
and  $d\tau'=\frac{dx}{x}\Delta\tau_d$ as an  approximation.
The integration $\int d\tau$ involving
$V(\tau)$, which has a factor of the form $e^{-a(\tau-\tau_d)^2}$.
As a stochastic quantity,   $\dot{h}(\tau)$
contains generally a mixture of oscillating modes,
such as
$e^{ik\tau}$ and $e^{-ik\tau}$,
and so does the spherical
Bessel function $j_l(k(\tau-\tau_0))$.
Thus $\dot{h}(\tau)   j_l(k(\tau-\tau_0)) $ generally contains
terms $\propto e^{-ibk(\tau-\tau_0)}$, where $b\in [-2, 2]$.
Using the formula
\[
\int_{-\infty}^{\infty}
e^{-ay^2}e^{ibky}dy=e^{-\frac{(bk)^2}{4a}}\int_{-\infty}^{\infty}
e^{-ay^2}dy,
\]
and  $V(\tau)$ in Eq. (\ref{halfgaussian1}) and (\ref{halfgaussian2}),
the integration  is approximated by
 \be
 \int_0^{\tau_0}d\tau V(\tau)
\dot{h}(\tau) j_l(k(\tau-\tau_0)) \approx
\frac{1}{2}  D(k)
\dot{h}(\tau_d)j_l(k(\tau_d-\tau_0)) \int_0^{\tau_0}d\tau V(\tau).
 \ee
where
 \be \label{D}
D(k)\equiv\frac{1}{2} [e^{-\alpha(k\Delta\tau_{d1})^2}
+e^{-\alpha(k\Delta\tau_{d2})^2}],
\ee
and  $\alpha$ takes values in the range  $[0,2]$,
depending on the phase of
$\dot{h}(\tau)   j_l(k(\tau-\tau_0)) $.
Here we take $\alpha$ as a parameter.
For the Gaussian  visibility function one
would have $D(k)\equiv e^{-\alpha(k\Delta\tau_d)^2}$.
The remaining integrations in $\beta_l$ is
\be
 \int_0^{\tau_0}d\tau V(\tau)
 \int_1^{\infty}\frac{dx}{x}e^{-\frac{3}{10}\kappa(\tau)
  x}e^{\frac{-7}{10}\kappa(\tau)} =
  \int_0^{\infty}d\kappa e^{-\frac{17}{10}\kappa}
 \int_1^{\infty}\frac{dx}{x}e^{-\frac{3}{10}\kappa x}
=  \frac{10}{17}\ln\frac{20}{3}. \ee
This number is the outcome from the second order of the
tight-coupling limit,
differing the first order
result $ \frac{10}{7}\ln\frac{10}{3}$ in Ref. \cite{prit}.
Finally one obtains
 \be \label{b}
 \beta_l(\tau_0)=
  \frac{1}{17}\ln\frac{20}{3} i^l
 \Delta\tau_d \dot{h}(\tau_d) j_l(k(\tau_d-\tau_0))D(k),
 \ee
which contains explicitly the time derivative
$\dot{h}(\tau_d)$ of RGW.
Substituting this back into Eqs. (\ref{gg}) and (\ref{cc}) yields
the polarization spectra
 \be\label{power}
 C^{XX}_{l}=\frac{1}{16\pi}\left(\frac{1}{17}\ln\frac{20}{3}\right)^2
 \int  P_{Xl}^2(k(\tau_d-\tau_0))|\dot{h}(\tau_d)|^2
\Delta\tau_{d}^2 D^2(k)\,k^2dk,
\ee
where ``X" denotes ``G" or ``C" the type of the CMB polarization.
For the electric type
 \be \label{pg}
 P_{Gl}(x)=\frac{(l+2)(l+1)}{(2l-1)(2l+1)}j_{l-2}(x)
 -\frac{6(l-1)(l+2)}{(2l-1)(2l+3)}j_{l}(x)
 +\frac{l(l-1)}{(2l+3)(2l+1)}j_{l+2}(x),
 \ee
 and for the magnetic type
 \be \label{pc}
 P_{Cl}(x)=\frac{2(l+2)}{2l+1}j_{l-1}(x)-\frac{2(l-1)}{2l+1}j_{l+1}(x).
 \ee
To completely determine $ C^{XX}_{l}$ above,
we need the initial amplitude $\dot{h}(\tau_d)$
to be fixed through Eq. (\ref{powerh})
by the initial spectrum $P_h(k)=rP_s$,
associated to the scalar spectrum $P_s$ by Eq. (\ref{ratio}).
WMAP observation\cite{map1-inflation} gives the scalar spectrum
\be
P_s(k_0)=2.95\times10^{-9}A(k_0),
\ee
with
$k_0=0.05$ Mpc$^{-1}$ and $A(k_0)=0.8$. Taking the
scale-invariant spectrum with $n_T= 0$ in (\ref{powerh}), then the
amplitude $A_T$ in (\ref{powerh}) depends on $r$.

\begin{figure}[pt]
\centerline{\includegraphics[width=12cm]{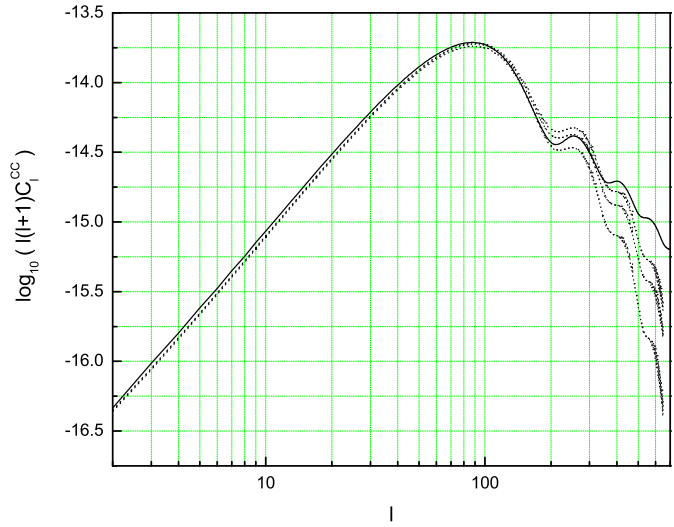}}
\caption{\label{magnetic}
The magnetic polarization spectrum $C^{CC}_l$ with the ratio $r=1$.
The solid line is the numerical result from the CMBFAST.
The upper dot line is the result from
the half-Gaussian visibility function with $\alpha=1.7$,
the middle dot line is  with $\alpha=2$,
and the lower dot line is the  Gaussian fitting with $\alpha=2$.
The half-Gaussian fitting is better than the Gaussian one.}
 \end{figure}

The polarization spectrum $C_{l}^{CC}$  of magnetic type,
calculated from our analytic formulae (\ref{power})
and from the numerical  CMBFAST,\cite{CMBFAST}
are shown in Fig. \ref{magnetic}.
The approximate analytic  result
is quite close to that of the numerical CMBFAST
for the first three peaks that are observable.

{\flushleft \emph{The location of the peaks:}}

\begin{figure}[pt]
\centerline{\includegraphics[width=12cm]{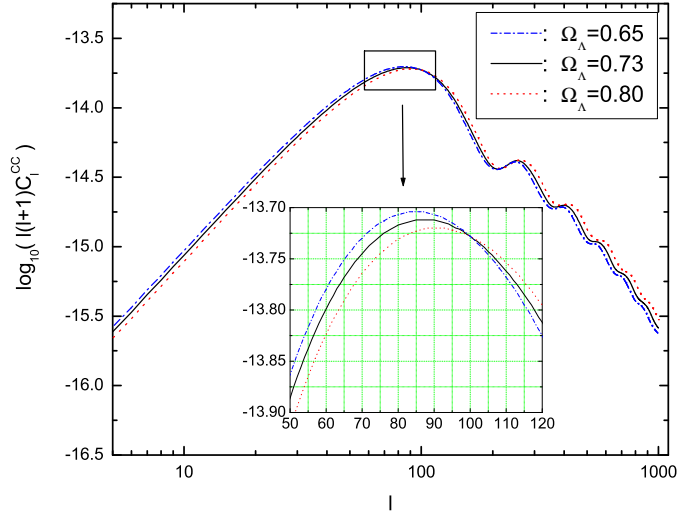}}
\caption{\label{ol}
$C_l^{CC}$ depends weakly on the dark energy.
A smaller  $\Omega_{\Lambda}$ yields a higher amplitude and
shifts the peaks to  larger scales.}
 \end{figure}

\begin{figure}[pt]
\centerline{\includegraphics[width=12cm]{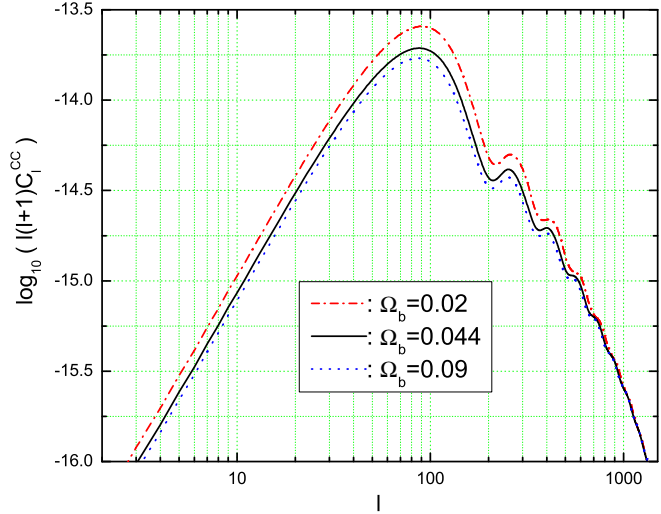}}
 \caption{\label{Omega-b}
 The dependence of $C_l^{CC}$ on
 $\Omega_b$ in the $\Lambda$CDM universe with
$\Omega_{\Lambda}=0.73$,
$\Omega_{dm}=1-\Omega_{\Lambda}-\Omega_b$, and $r=1$.
A larger $\Omega_b$ yields a lower amplitude and
shifts the peaks slightly to  larger scales.}
 \end{figure}

In (\ref{pg}) and  (\ref{pc})
the spherical Bessel function $j_l(k(\tau_d-\tau_0))$ is
peaked at $l\simeq k(\tau_0-\tau_d)\simeq k\tau_0$ for $l\gg1$.
So the peak location of
the power spectra are directly determined by
 \be \label{c}
 C_l^{XX}\propto
 \left|\dot{h}(\tau_d)\right| ^2k^2D^2(k)\left.\right|_{k=l/\tau_0}.
 \ee
The factor $D(k)$ has a larger damping at larger $l$,
so the first peak of the
power spectrum has the highest amplitude.
From our analytic solution one has
$\dot{h}(\tau_d)^2\propto [j_2(k\tau_d)]^2$,
which peaks at $k\tau_d\simeq3$.
 Thus $C_l^{XX}$ peaks around
 \be \label{l}
l\simeq k\tau_0\simeq3\tau_0/\tau_d.
 \ee
A lower dark energy component leads to smaller $\tau_0/\tau_d$.
For $\Omega_{\Lambda}=0.65$,
$0.73$ and $0.80$, respectively, and with fixed $\Omega_b=0.044$,
$\Omega_{dm}=1-\Omega_{\Lambda}-\Omega_b$,
a numerical calculation yields that
  $\tau_0/\tau_d\simeq50.1$, $51.3$ and $53.6$, respectively.
A smaller $\Omega_{\Lambda}$ will also shift the
peaks  $\dot{h}(\tau_d)$  slightly to larger scales.
Together,
 a smaller $\Omega_{\Lambda}$ will shift the
peak of  $C_l^{XX}$ to larger scales, as demonstrated in Fig. \ref{ol}.
This suggests a new
way to study the cosmic dark energy.
The baryon component also influence the peak location.
A higher baryon  density $\Omega_b$ makes the peak to
large scales, as  is demonstrated  in Fig. \ref{Omega-b}.

{\flushleft \emph{ The height of amplitude:}}

 $C_l^{XX}$  depend
on the decoupling  thickness $\Delta\tau_d$
and the damping factor $D(k)$:
$C_l^{XX}\propto \Delta\tau_d^2 D(k)^2$.
For a fixed $k$, the smaller $\Delta\tau_d$
leads to a larger $D(k)$.
$\Delta\tau_d$ is mainly
determined by the baryon density $\Omega_b$ of the Universe.
A higher $\Omega_b$ corresponds to a
smaller $\Delta\tau_d$.
The total effect is that   a higher $\Omega_b$ leads to a lower  $C_{l}^{CC}$,
as is shown in Fig. \ref{Omega-b}.
Besides, a smaller $\Omega_{\Lambda}$ yields a higher amplitude,
as seen in  Fig. \ref{ol}.

{\flushleft \emph{The influence of the inflation on  $C_l^{XX}$:}}

The inflation models determine the ratio $r$ and
the amplitude $|\dot{h}(\tau_d)|$ in (\ref{powerh}),
including the spectrum index $n_T$ of RGW.
A larger $r$ yields a larger $|\dot{h}(\tau_d)|$  and a higher
polarization.
A larger $n_T$ yields
higher polarization spectra.

The more recent treatments can be found
in Refs. \cite{zhaozhang083006},
\cite{baskaran} $-$ \cite{Grishchuk07}.

\section{ Conclusion and Discussion}

Our calculations of RGW have shown that
in the low frequency range  the peak of spectrum is now
located at a frequency $\nu_E \simeq
(\frac{\Omega_m}{\Omega_{\Lambda}})^{1/3} \nu_H$, where $\nu_H$ is
the Hubble frequency, and there appears a new segment of spectrum
between  $\nu_E$ and $\nu_H$. In all other intervals of
frequencies $\geq  \nu_H$, the spectral amplitude acquires an
extra factor $\frac{\Omega_m}{\Omega_{\Lambda}}$, due to the
current acceleration, otherwise the shape of spectrum is similar
to that in the decelerating models.
 The amplitude  for the model $\Omega_{\Lambda}=0.65$ is
$\sim 50 \%$ greater than that of the model $\Omega_{\Lambda}=0.7$.
The spectrum sensitively depends
on the inflationary models, and a larger $\beta$ yields  a flatter
spectrum, producing  more power.
Both the LIGO bound  and the
nucleosynthesis bound point out that the
inflationary model $\beta=-1.8$ is
ruled out, but the model $\beta=-2.0$ is still alive.

Our  analytic polarization spectra of CMB has the
following improvements.

(i)  The analytic  result of CMB polarization is quite
close to the numerical result from the CMBFAST code.
The dependence of polarization on
the dark energy and the baryons are analyzed.
A smaller $\Omega_{\Lambda}$ yields a higher amplitude
and shifts the peaks to large scales.
A larger $\Omega_b$ yields a lower amplitude
and shifts the peaks to large scales.

(ii) Our half-Gaussian approximation of the visibility function
fits analytically better than the simple Gaussian fitting,
and its time integration yields a
parameter-dependent damping factor.
This improves the spectrum $ \sim 30\%  $ around the second
and third peaks.

(iii) The second order of tight coupling limit
reduces the amplitude of spectra by $\sim 58\%$,
comparing with the first order.

(iv)  A larger value of the spectrum index $n_T$  of RGW
and a larger ratio $r$ yield higher polarization spectra.

{\flushleft ACKNOWLEDGMENT:
Y.  Zhang would like to thank the organizers of
Third International ASTROD Symposium
on Laser Astrodynamics, Space Test of Relativity
and Gravitational-Wave Astronomy.
He also thanks Dr. L. Q. Wen for interesting discussions.
The work has been supported by the CNSF No. 10773009, SRFDP,  and CAS.
W. Zhao has been supported by Graduate Student Research
Funding from USTC.}


\begin{thebibliography}{0}


\bibitem{starobinsky}
  V. A. Rubakov, M.V. Sazhin and A.V. Veryaskin,
              \emph{Phys. Lett. B} {\bf  115} (1982) 189.

\bibitem{fabbris}  R. Fabbri and M. D. Pollock,
              \emph{Phys. Lett. B} {\bf 125} (1983) 445.

\bibitem{AbbottWise}  L. Abbott and M. Wise,
              \emph{Nuc. Phys. B}  {\bf  237} (1984) 226.

\bibitem{AbbottHarari}  L. F. Abbott and D. D. Harari,
              \emph{Nucl. Phys. B} {\bf  264}  (1986) 487.

\bibitem{Allen}  B. Allen,
             \emph{ Phys. Rev. D} {\bf 37}  (1988) 2078.

\bibitem{Sahni}  V. Sahni,    \emph{Phys. Rev. D} {\bf 42}  (1990) 453.

\bibitem{Grishchuk77}  L. P. Grishchuk,
             \emph{Ann. NY Acad. Sci.} {\bf 302}  (1977) 439.

\bibitem{Basko} M. M. Basko and A. G. Polnarev,
         \emph{ Mon. Not. R. Astron. Soc.} {\bf 191} (1980) 207.

\bibitem{kaiser} N. Kaiser,
         \emph{ Mon. Not. R. Astron. Soc.}  {\bf 202} (1983)  1169;

\bibitem{BondEfstathiou84}   J.R. Bond and G. Efstathiou,
              \emph{Astrophys. J. Lett.}  {\bf 285} (1984)  L45;

\bibitem{BondEfstathiou87}   J.R. Bond and G. Efstathiou,
              \emph{Mon. Not. R. Astron. Soc.} {\bf 226} (1987)  655.

\bibitem{Polnarev80}
   M. M. Basko and A. G. Polnarev,
           \emph{Mon. Not. R. Astron. Soc.} {\bf 191} (1980)  207.

\bibitem{Polnarev85}   A. Polnarev, \emph{Sov. Astron.}  {\bf 29} (1985) 6.

\bibitem{Frewin94}   R. A. Frewin, A.G. Polnarev and P. Coles,
                \emph{Mon. Not. R. Astron. Soc.}  {\bf 266} (1994) L21.

\bibitem{Keating98}   B. Keating  {\it et al}, \emph{Astrophys. J.} {\bf 495} (1998) 580.

\bibitem{zalda}
          D. Harari and M. Zaldarriaga,
                   \emph{Phys. Lett.  B} {\bf  310} (1993) 96.

\bibitem{Zaldarriaga Harari}   M. Zaldarriaga and D. D. Harari,
                   \emph{ Phys. Rev. D} {\bf  52}  (1995) 3276.

\bibitem{Ng}  K. L. Ng and K. W. Ng,
          \emph{Astrophys. J.} {\bf 445} (1995) 521.

\bibitem{kosowsky}
         A. Kosowsky, \emph{Ann. Phys.} {\bf 246} (1996) 49.

\bibitem{Kamionkowski97}  M. Kamionkowski, A. Kosowsky and  A. Stebbins,
                    \emph{Phys. Rev. D}  {\bf 55} (1997) 7368.

\bibitem{Crittenden93}   R. Crittenden, R. L. Davis and P. J. Steinhardt,
                    \emph{Astrophys. J.} {\bf 417} (1993) L13.

\bibitem{Crittenden94}   D. Coulson, R. Crittenden and N. Turok,
                    \emph{Phys. Rev. Lett.} {\bf 73} (1994) 2390.

\bibitem{Crittenden95}   R. Crittenden,  D. Coulson,  and N. Turok,
                    \emph{Phys. Rev. D} {\bf 52} (1995) 5402.

\bibitem{SeljakZaldarriaga}   U. Seljak and M. Zaldarriaga,
                  \emph{Phys. Rev. Lett.} {\bf28} (1997) 2054.

\bibitem{HuWhite}         W. Hu and M. White,
                  \emph{Phys. Rev. D} {\bf  56} (1997) 597.

\bibitem{Ni} W. T. Ni,
            \emph{Chin. Phys. Lett.} {\bf 22} (2005) 33.

\bibitem{Ni2}  W. T. Ni, \emph{Int. J. Mod. Phys. D}  {\bf  14}  (2005)  901.

\bibitem{grish1} L. Grishchuk,
             \emph{Class. Quant. Grav.} {\bf 14}  (1997) 1445.

\bibitem{grish01} L. Grishchuk,  Lecture Notes Physics {\bf 562}, (2001) 164.

\bibitem{zh}  Y. Zhang  {\it et al},
                 \emph{Class. Quant. Grav.} {\bf 22} (2005) 1383 .

\bibitem{zh05}  Y. Zhang and  W. Zhao,
                \emph{Chin. Phys. Lett.} {\bf 22}   (2005) 1817.

\bibitem{zh06}   Y. Zhang {\it et al},
              \emph{Class. Quant. Grav.} {\bf 23}  (2006) 3783.

\bibitem{spergel} D. N. Spergel {\it et al},
               \emph{Astrophys. J. Suppl.} {\bf 148}  (2003)  175.

\bibitem{spergel2}   D. N. Spergel {\it et al},
               \emph{Astrophys. J. Suppl.} {\bf 170} (2007) 377.

\bibitem{ligo} http://www.ligo.caltech.edu/advLIGO.


\bibitem{BAbbbottprl}  B. Abbott {\it et al},
            \emph{Phys. Rev. Lett.} {\bf94},  (2005) 181103.

\bibitem{BAbbbott05}  B. Abbott {\it et al},
            \emph{Phys. Rev. Lett.} {\bf 95}  (2005) 221101.

\bibitem{maggiore} M. Maggiore,
        \emph{Phys. Rep. } {\bf331} (2000) 283.

\bibitem{maia}  M. R. G. Maia,
               \emph{ Phys. Rev. D} {\bf  48}  (1993) 647.

\bibitem{maia2} M. R. G. Maia and J.D. Barrow,
        \emph{Phys. Rev. D} {\bf  50}  (1994) 6262.

\bibitem{Tashiro} H. Tashiro, K. Chiba and M. Sasaki,
          \emph{Class. Quant. Grav.} {\bf 21}   (2004) 1761.

\bibitem{Henriques} A. B. Henriques,
         \emph{Class. Quant. Grav.}  {\bf 21}   (2004) 3057.

\bibitem{Gong} G. Gong,
         \emph{Class. Quant. Grav.} {\bf21} (2004) 5555.

\bibitem{Zhao043503} W. Zhao and Y. Zhang,
      \emph{Phys. Rev. D} {\bf 74}  (2006) 043503.

\bibitem{WangZhang08} S. Wang, Y. Zhang, T. Y. Xia  and H. X. Miao,
       accepted, \emph{Phys. Rev. D} {\bf 77}  (2008) 104016.

\bibitem{weinberg} S.  Weinberg,
      \emph{ Phys. Rev. D} {\bf 69}  (2004) 023503.

\bibitem{DicusRepko} D. A. Dicus and  W. W. Repko,
          \emph{Phys. Rev. D} {\bf 72}  (2005) 088302.

\bibitem{WatanabeKomatsu} Y. Watanabe and E. Komatsu,
    \emph{Phys. Rev. D} {\bf  73}   (2006) 123515.

\bibitem{MiaoZhang} H. X. Miao and Y. Zhang,
           \emph{Phys. Rev. D} {\bf  75}  (2007) 104009.



\bibitem{chan} S. Chandrasekhar,
               \emph{Radiative Transfer},  Dover, New York (1960).

\bibitem{zhz}   Y. Zhang, H. Hao and W. Zhao,
             \emph{ ChA\&A} {\bf 29} (2005) 250.

\bibitem{peebles} P.  J. E. Peebles,
                 \emph{Astrophys. J.} {\bf 153} (1968) 1.

\bibitem{JonesWyse}  B. Jones and R. Wyse,
                    \emph{Astron. Astrophys.}  {\bf 149}  (1985) 144.

\bibitem{Hu}  W. Hu and N. Sugiyama,
           \emph{Astrophys. J.} {\bf 444} (1995) 489.

\bibitem{prit}
         J. R. Pritchard and M. Kamionkowski,
                   \emph{Ann. Phys.} {\bf 318} (2005) 2.

\bibitem{zhaozhang083006}  W. Zhao and Y. Zhang,
                 \emph{Phys. Rev. D} { \bf 74} (2006) 083006.

\bibitem{map1-inflation} H. V. Peiris {\it et al},
           \emph{Astrophys. J. Suppl.} {\bf 148} (2003) 213.

\bibitem{CMBFAST}
       U. Seljak and M. Zaldarriaga,
         \emph{Astrophys. J.} {\bf 469} (1996) 437.

\bibitem{baskaran}  D. Baskaran, L. Grishchuk and A. Polnarev,
                 \emph{Mon. Not. R. Astron. Soc.} {\bf 370} (2006)  799.

\bibitem{KeatingPolnarev}  B. Keating, A. Polnarev, N. Miller and D. Baskaran,
                 \emph{Int. J. Mod. Phys. B }{\bf  21} (2006) 2459.

\bibitem{Polnarev07}  A. Polnarev, N. Miller and  B. Keating,
                    arXiv:0710.3649 astro-ph.

\bibitem{Grishchuk07}  L. P. Grishchuk, arXiv:0707.3319 astro-ph.
\end{thebibliography}
\end{document}